\documentclass[12pt,a4paper,final]{iopart}

\usepackage{iopams}  
\usepackage{graphicx}
\usepackage[breaklinks=true,colorlinks=true,linkcolor=blue,urlcolor=blue,citecolor=blue]{hyperref}

\newcommand{\be}{\begin{eqnarray}}
\newcommand{\ee}{\end{eqnarray}}

\begin{document}

\title[Flat-band many-body localization]
{Flat-band many-body localization and ergodicity breaking in the Creutz ladder}

\author{Yoshihito Kuno$^{1}$$^\ast$, Takahiro Orito$^{2}$$^\ast$ and Ikuo Ichinose$^{2}$}
\address{$^1$ Department of Physics, Graduate School of Science, Kyoto University, Kyoto 606-8502, Japan}
\address{$^2$ Department of Applied Physics, Nagoya Institute of Technology, Nagoya 466-8555, Japan}
\address{$^\ast$ These authors contributed equally to this work}
\ead{ykuno@yagura.scphys.kyoto-u.ac.jp}



\begin{abstract}
We study disorder-free many-body localization in
the flat-band Creutz ladder, which was recently realized in cold-atoms 
in an optical lattice.
In a non-interacting case, the flat-band structure of the system leads to a Wannier
wavefunction localized on four adjacent lattice sites. 
In the flat-band regime both with and without interactions, 
the level spacing analysis exhibits Poisson-like distribution
indicating the existence of disorder-free localization. 
Calculations of the inverse participation ratio support this observation.
Interestingly, this type of localization is robust to weak disorders, whereas for
strong disorders, the system exhibits a crossover into the conventional disorder-induced
many-body localizated phase. 
Physical picture of this crossover is investigated in detail.
We also observe non-ergodic dynamics in the flat-band regime without disorder.
The memory of an initial density wave pattern is preserved for long times.
\end{abstract}

\vspace{2pc}
\maketitle

\section{Introduction}\label{intro}

Localization in non-interacting electron systems has been
extensively studied since Anderson discussed the disorder effect on the single-particle 
electron wavefunction in solids \cite{Anderson}. 
Presently, what is called Anderson localization (AL), is recognized as a universal phenomenon
in various physical systems \cite{Lagendijk}. 
In AL quantum system, a single-particle electron wavefunction is exponentially localized
with a finite localization length, and an insulating phase forms. 
Owing to the recent development in the computational power and numerical techniques, 
study on the effect of the interactions between particles on AL is currently
one of the main research topics in condensed matter physics. 
It is now recognized that AL persists in some cases even if the particles interact.
This is called many-body localization (MBL). 
Mostly by numerical simulations, it has been clarified that the MBL phase exhibits
some characteristic properties such as Poisson distribution in the level spacing analysis
(LSA) of the energy eigenvalues similar to that of the conventional AL and 
the logarithmic growth of entanglement entropy.
In its glassy dynamics,
MBL is closely related with the breaking of eigenstate thermalization hypothesis and 
ergodicity breaking dynamics \cite{Nandkishore,Basko,Abanin,Alet,Oganesyan,Abanin2}. 
This means that a closed ergodicity-breaking system does not thermalize for a long time,
and if we prepare a non-entangled initial state in such a system, 
the information of the intial state is conserved for a long time without being lost.
Recent experiments on cold-atom gases in optical lattices have reported 
evidences for the existence of MBL phenomena \cite{Schreiber,Choi,Lukin,Rispoli}.

Until recently, most of the theoretical studies have focused on MBL
induced by the disorders encoded in the on-site potentials, hopping amplitudes 
and interactions, as well as quasi-periodic potentials \cite{Iyer}. 
On the other hand very recently, disorder-free  AL/MBL-like phenomena have
been revealed in a Wannier--Stark ladder \cite{Nieuwenburg,Schulz,Orito}, 
dipolar atom gases in an optical lattice \cite{Li}, some lattice-gauge theoretical models \cite{Smith1,Smith2,Brenes,Takaishi}, quantum Hall systems \cite{Krishna}, 
a diamond chain system \cite{vidal1,vidal2,Cartwright}, and a disorder-free spin chain \cite{Kormos,Mazza,Liu,Lerose}. 

Motivated by the above findings, we shall report another type of disorder-free 
MBL system in this paper. 
It is a flat-band system with interactions. 
Certain flat-band structure suppresses particle hoppings effectively 
and generates a localized Wannier state \cite{Takayoshi,Mondaini} that is similar to the 
localized states in the conventional AL system. 
Such a localized Wannier state was theoretically investigated for certain
{\em non-interacting flat-band systems} with and without weak disorders \cite{Flach,Leykam}.
We are motivated by the existence of such localized wavefunctions and 
study a flat-band type localization in the Creutz ladder \cite{Creutz}.
The Creutz ladder is a simple model and also experimentally feasible 
in cold atom gases. So far, there are several theoretical proposals for implementation of the model \cite{Bermudez5,Bermudez6,Bermudez2,ours1}, and 
cold atom experiments realized some related systems \cite{synthetic_dim1,synthetic_dim2,Kang},
whose the physical properties have been extensively 
studied  \cite{Bermudez1,Bermudez3,Bermudez4}.

This paper is organized as follows.
In Sec.~\ref{model}, we introduce the target Creutz ladder model.
We focus on the flat-band case, and analytically study properties of the flat-band
states.
We explicitly reveal the origin of localization and discuss the possibility of MBL
with repulsions.
Effects of on-site disorders are also discussed, and the global phase structure is given.

In Sec.~\ref{results}, we present results of the numerical study.
We first perform the LSA  and also the level-spacing-ratio (LSR) analysis for the system
with weak disorders under the flat-band condition and
find that the probability distribution exhibits Poissonian behavior
for both the non-interacting and interacting cases, 
indicating a localization tendency. 
Interestingly enough as the strength of the disorder is increased, 
we find that both the LSA and LSR exhibit behavior of
Gaussian unitary ensemble (GUE) corresponding to extended (delocalized) states.
These results are compared with those of the non-flat case in order to 
clarify the difference between the flat-band and non-flat band cases.
The above phenomenon is discussed via the analytical study in Sec.~\ref{model}.
Then, we investigate the inverse participation ratio (IPR) to find that its results
corroborate the localization tendency of the flat-band Creutz model. 
In particular, energy-resolved IPR exhibits very interesting behavior, which
explicitly clarifies typical properties of the flat-band states as increasing
the strength of disorder.
We finally investigate distribution of the localization length for typical disorder strengths.
Energy-resolved distribution reveals origin of the crossover observed by 
the LSA and IPR.

In the final subsection of Sec.~\ref{results}, we study the dynamics in 
the flat-band Creutz ladder, i.e., 
we investigate the time-evolution of states in which fermions are periodically put on sites.
The result shows ergodicity-breaking dynamics, i.e., the memory of the particle distribution
in initial states is preserved for long times.  
Besides the above important result of the non-ergodicity of the Creutz ladder,
we find another interesting phenomenon for the cases of $1/6$ and $1/4$-particle filling.

Section \ref{conclusion} is devoted for conclusion.
We present the summary and also give future perspective.

\section{Creutz ladder model and flat-band localization}\label{model}

\begin{figure}[t]
\centering
\includegraphics[width=8cm]{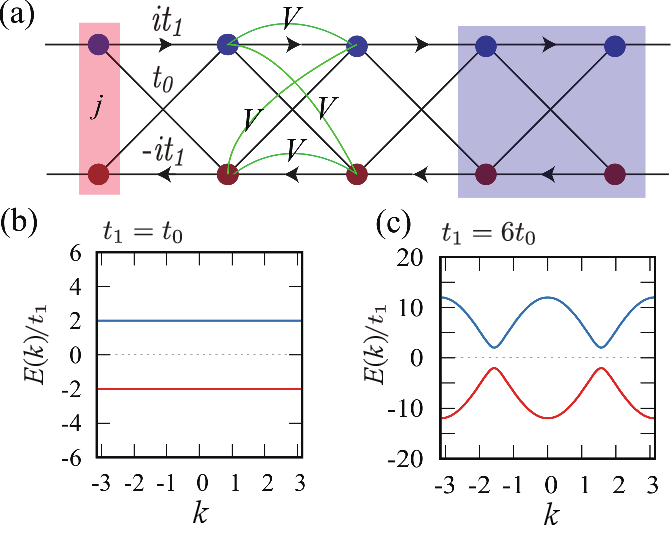}
\caption{(a) Creutz ladder: the red shaded area represents a unit cell and the blue one is a Wannier state under the flat-band condition. 
(b) Flat-band structure. (c) Non-flat-band band structure.}
\label{Setup}
\end{figure}

In this work, we study an interacting Creutz ladder model with 
the Hamiltonian \cite{Creutz}, 
\begin{eqnarray}
H&=&\sum_{j}\biggl[-it_1(a^{\dagger}_{j+1}a_{j}-b^{\dagger}_{j+1}b_{j})
-t_{0}(a^{\dagger}_{j+1}b_{j}+b^{\dagger}_{j+1}a_{j}) +\mbox{h.c.}  \nonumber\\
&+&V(n_{a,j}n_{a,j+1}+n_{b,j}n_{b,j+1}+n_{a,j}n_{b,j+1}+n_{b,j}n_{a,j+1})\nonumber\\
&+&\mu_{a,j}n_{a,j}+\mu_{b,j}n_{b,j}\biggr],
\label{Creutz}
\end{eqnarray}
where  $a^{(\dagger)}_{j}$ and $b^{(\dagger)}_{j}$ are the fermion annihilation (creation) operators on the upper and lower chains, respectively, and 
subscript $j$ denotes a unit cell. 
$n_{a(b),j}$ is the number operator of the particle on the upper (lower) chain. 
$t_1$ and $t_{0}$ are the intra-chain and inter-chain hopping amplitudes, respectively.
$V$ is the intra-chain and inter-chain repulsions, as depicted in Fig.~\ref{Setup} (a), 
which is one of the simplest interactions suitable for the present study as we explain
shortly. 
There are two possible ways to implement this type of interactions in real experiments:
(I) Method to use electric or magnetic dipole-dipole interactions between atoms \cite{Lahaye, Baier}, 
(II) To use natural overlap of wannier functions between neighboring sites connected by 
horizontal and diagonal links induces to this type of interaction. 
The case (I) may induce vertical interactions, but we ignore them in this work. 
We verified that the vertical interactions do not change the subsequent numerical results substantially.
Obviously, the repulsive $V$-interaction prefers the density-wave configurations
in the ladder direction. 
$\mu_{a(b),j}$ is a random disorder chemical potential, which has a uniform
distribution, such as $\mu_{a(b),j}\in [-\mu/2,\mu/2]$, and breaks the chiral symmetry \cite{diss_come}.
This choice of the disorder plays a significant role in the localization phenomenon
in the present model as we explain shortly.

The energy spectrum of the non-interacting case of $H$ in Eq.~(\ref{Creutz}), 
with $V=\mu=0$ is given as $E(k)=\pm\sqrt{(2t_1\sin k)^2+(2t_{0}\cos k)^{2}}$, 
where $k$ is the wave number and the bandwidth is $|2(t_{1}-t_0)|$. 
As shown in Figs.~\ref{Setup} (b) and (c), the band is flat for $t_0=t_1$ with
$E(k)=\mp 2t_0$, whereas it is dispersive for $t_{0}\neq t_{1}$ \cite{t16t0}. 
The non-interacting case of $H$ in Eq.~(\ref{Creutz}) with $\mu=0$ belongs to 
the {\rm BDI} class in the topological classification theory \cite{Schnyder,Kitaev,Velasco,Gholizadeh}.
Hence, the model has chiral, time-reversal and particle-hole symmetries.
In particular, the chiral symmetry makes the energy spectrum symmetric 
around zero energy.
In addition, at the flat-band point, $t_0=t_{1}$, a localized Wannier state exists in the system,
whose wavefunction for the lower spectrum is given by \cite{Takayoshi,Mondaini} 
\begin{eqnarray}
|\Psi_w\rangle_j = -\frac{1}{2}\biggl[a^{\dagger}_{j}+ia^{\dagger}_{j+1}+b^\dagger_{j}
-ib^{\dagger}_{j+1}\biggr] |0\rangle,
\label{localized_gw}
\end{eqnarray}
where $|0\rangle$ is the empty state. 
The state $|\Psi_w\rangle_j$ spans over two adjacent unit cells, i.e., 
it is a four-site localized state, and there are two $|\Psi_w\rangle_j$'s per site.

It is quite useful to study analytically the flat-band case of the present system 
for the forthcoming numerical investigation.
In that case, the hopping part of the Hamiltonian reduces to the following one, 
$H_{\rm flat}$,
\begin{eqnarray}
H_{\rm flat}&=&\sum_j\Big[
-it_0(a^\dagger_{j+1}a_j-b^\dagger_{j+1}b_j)-t_0(a^\dagger_{j+1}b_j+b^\dagger_{j+1}a_j) 
+\mbox{h.c.}\Big].
\label{Hflat}
\end{eqnarray}
Then, we introduce the following operators,
\begin{eqnarray}
w_{Aj}=a_j+ib_j,  \;\; w_{Bj}=a_j-ib_j,
\label{ww}
\end{eqnarray}
where we can prove $\{ w^\dagger_{Aj}, w_{Bj} \}=0$. 
This transformation is a kind of detangling for a lattice system \cite{Flach}. 
Under this transformation, the Creutz ladder is detangled into a simple lattice system where each lattice site is completely decoupled each other.
In terms of $w_{Aj}$ and $w_{Bj}$, $H_{\rm flat}$ is expressed as, 
\begin{eqnarray}
H_{\rm flat}=\sum_j\Big[-it_0w^\dagger_{A,j+1}w_{Bj}
+it_0w^\dagger_{Bj}w_{A,j+1}\Big],
\label{Hflat2}
\end{eqnarray}
and straightforward manipulations show,
\begin{eqnarray}
H_{\rm flat}w^\dagger_{Aj}|0\rangle=2it_0 w^\dagger_{B,j-1}|0\rangle,\;\;\;\; 
H_{\rm flat}w^\dagger_{Bj}|0\rangle=-2it_0 w^\dagger_{A,j+1}|0\rangle.
\label{switch}
\end{eqnarray}
Equations in Eq.~(\ref{switch}) reveal very important properties of the Creutz 
ladder mode with the flat-band coupling, i.e., in terms of $\{ w_A, w_B\}$-`particles', 
$w_{A(B)}$-particle hops only left (right)-hand site and changes to 
$w_{B(A)}$-particle.
Therefore, the $\{ w_A, w_B\}$-particles strictly localize on two adjacent
rungs of the ladder.
It is obvious that the Wannier state in Eq.~(\ref{localized_gw}) is nothing but 
a static state composed of a pair of nearest-neighbor $\{ w_A, w_B\}$ such as 
$$
|\Psi_w\rangle_j =-{1\over 2}(iw^\dagger_{A,j+1}+w^\dagger_{Bj})|0\rangle.
$$
Similarly, the upper-spectrum eigenstates can be constructed easily as 
$
(iw^\dagger_{A,j+1}-w^\dagger_{Bj})|0\rangle. 
$
Therefore, the flat-band Hamiltonian, $H_{\rm flat}$, can be expressed in terms of
the following operators, $W^{\pm\dagger}_j$, that create energy eigenstates,
\begin{eqnarray}
&&H_{\rm flat}=\sum_{j}\Big[-2t_0W^{+\dagger}_{j}W^{+}_{j}
+2t_0W^{-\dagger}_{j}W^{-}_{j}\Big],   \nonumber   \\
&&W^{\pm\dagger}_j\equiv {1 \over 2} (iw^\dagger_{A,j+1}\pm w^\dagger_{B,j}).
\label{Hflat3}
\end{eqnarray}

One may wonder how the original fermion $a_j (b_j)$ behaves.
Obviously, they do {\em not} create an eigenstate of the Hamiltonian.
However, $a_j$ and $b_j$ are a simple superposition of $w_{Aj}$ and $w_{Bj}$,
i.e., $a^\dagger_j={1 \over 2}(w^\dagger_{Aj}+w^\dagger_{Bj})$.
Then, the time evolution of the state 
$a^\dagger_j|0\rangle={1 \over 2}(w^\dagger_{Aj}+w^\dagger_{Bj})|0\rangle$ 
can be easily obtained.
In fact as $w_{A(B)}$-particle hops only left (right)-hand site and changes to 
$w_{B(A)}$-particle, the resultant state of the time evolution 
is a superposition of the two states $a^\dagger_j|0\rangle$ and 
$(w^\dagger_{A,j+1}-w^\dagger_{B,j-1})|0\rangle$.
By straightforward calculations, we have,
\begin{eqnarray}
&&e^{-i\frac{H_{\rm flat}}{\hbar}t}w^\dagger_{Aj}|0\rangle 
=\cos\biggl(\frac{2t_0}{\hbar}t\biggr) w^\dagger_{Aj}|0\rangle - \sin\biggl(\frac{2t_0}{\hbar}t\biggr) w^\dagger_{Bj-1}|0\rangle,  \nonumber \\
&&e^{-i\frac{H_{\rm flat}}{\hbar}t}w^\dagger_{Bj}|0\rangle 
=\cos\biggl(\frac{2t_0}{\hbar}t\biggr) w^\dagger_{Bj}|0\rangle + \sin\biggl(\frac{2t_0}{\hbar}t\biggr) w^\dagger_{Aj+1}|0\rangle,
\label{uni1}
\end{eqnarray}
and therefore, the dynamics of the state $|\psi_{\rm ini}\rangle = a^{\dagger}_{j}|0\rangle$
is given by
\begin{eqnarray}
|\Psi_a(t)\rangle &\equiv& e^{-i\frac{H_{\rm flat}}{\hbar}t}|\psi_{\rm ini}\rangle
\nonumber  \\
&&=\cos\biggl(\frac{2t_0}{\hbar}t\biggr) a^{\dagger}_{j}|0\rangle + \sin\biggl(\frac{2t_0}{\hbar}t\biggr)
{1\over 2}(w^\dagger_{A,j+1}-w^{\dagger}_{B,j-1})|0\rangle.  \label{solution}
\end{eqnarray}
The above state in Eq.~(\ref{solution}) is obviously localized.

The analytical study in the above gives the following important observations 
on the Creutz ladder model in Eq.~(\ref{Creutz});
\begin{enumerate}
\item In the clean and non-interacting flat-band case, the Creutz ladder system 
is strictly non-ergodic and all eigenstates are localized.
\item The localization `length' is four lattice sites.
The Wannier state in Eq.~(\ref{localized_gw}) resides on four sites.
In the state expressed by Eq.~(\ref{solution}), a particle resides on a single site
and six sites with equal probability. 
Such a localized particle can be regarded as a concrete example of a flat-band compactons. 
More general discussion and construction for the flat band compactons has been given in \cite{VicencioFB,Morales-Inostroza}.
\item Under a disorder such as $\mu_{a,j}=\mu_{b,j}$, the $w$-particle picture is
robust, i.e., no on-site mixing of the $w_A$ and $w_B$-particles takes place, 
and therefore the above localization properties are intact.
On the other hand, a disorder such as $\mu_{a,j}\neq\mu_{b,j}$, which we employ in the
present work, tends to break the $w$-particle picture as it induces an on-site mixing.
\item Similarly, the interaction term in Eq.~(\ref{Creutz}) is expressed by 
the $w$-particle in the diagonal form,
$$
V\sum_j (w^\dagger_{Aj}w_{Aj}+w^\dagger_{Bj}w_{Bj})
(w^\dagger_{A,j+1}w_{A,j+1}+w^\dagger_{B,j+1}w_{B,j+1}),
$$
and therefore, the $w$-particle picture is robust even in the presence of the interaction.
\end{enumerate}

Before going into the practical calculations, we shall give some comments.
(1) In the following section, we consider the $1/8$-filling case. 
In such a low commensurate filling, 
particles described by Eq.~(\ref{localized_gw}) do not overlap substantially~\cite{Huber}. 
Then, it is expected that the $w$-particle picture is preserved even for rather strong 
$V$-interactions under {\em weak disorder}, and the system exhibits localization.
{\em This is nothing but a new kind of MBL}. 
The conventional disorder-induced MBL needs sufficiently strong disorders \cite{Alet}. 
On the other hand, our considering MBL is induced by the
flat band, i.e., distractive interference of hoppings.
(2) In the ordinary AL systems, localization length depends on the disorder strength.
On the other hand in the above MBL regime, the Wannier state in the flat-band has finite
components in {\em definite lattice sites}. 
We note that this properties give certain suggestions on 
the set up of an initial state for observing MBL dynamics in simulations
that we shall give in later section. 
(3) Increasing the disorder strength $\mu$, the $w$-particle picture is getting unstable,
and the genuine flat-band localization is expected to be destroyed. 
We expect that a crossover takes place from the flat-band localized states to 
a new kind of states at a critical disorder strength, $\mu_c$.

\section{Numerical studies}\label{results}

In this section, we shall study the Creutz model by the numerical methods.
As a hallmark of localization and (non)ergodicity, we investigate the level spacing, 
the inverse participation ratio and the temporal evolution of inhomogeneous states.
Obtained results all support the picture of the flat-band localization given
in Sec.~\ref{model}.
Furthermore, the numerical results show interesting behavior of the model,
in particular at
relatively high fillings, which come from the interplay between the locality of 
the flat-band regime and the repulsion. 
In what follows, we employ $t_{1}$ as a unit of energy.

\subsection{Level spacing analysis}\label{LSA}

\begin{figure}[h]
\centering
\includegraphics[width=8cm]{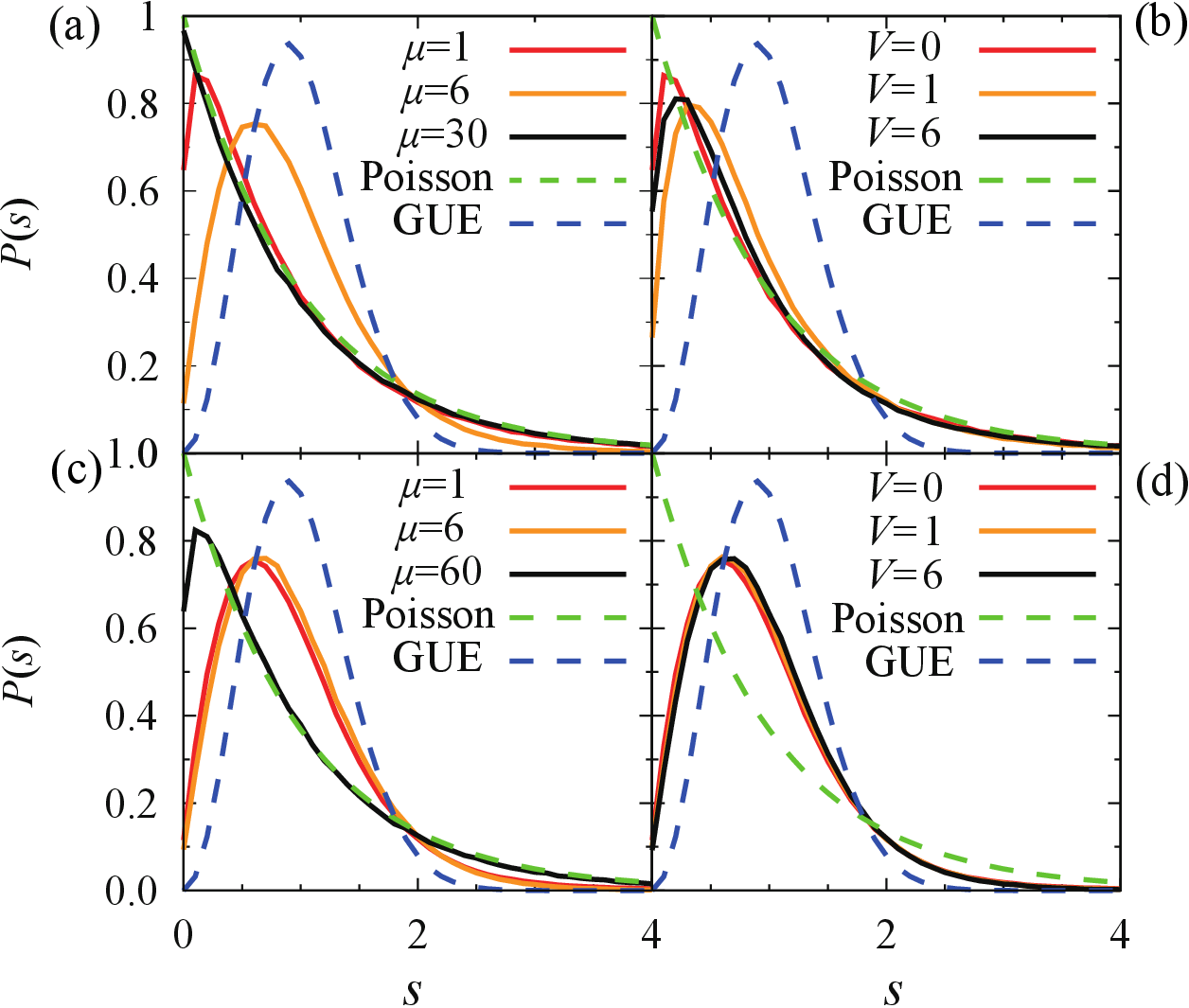}
\caption{Level spacing analysis: (a) Disorder dependence in non-interacting flat-band. 
(b) Interaction dependence in disordered flat-band ($\mu=1$). (c) Disorder dependence in non-interacting non-flat-band ($V=0$).
(d) Interaction dependence in disordered non-flat-band ($\mu=1$). 
For all cases, we employed $L=16$ and $N=4$ and averaged over 20 disorder realizations.}
\label{Fig2}
\end{figure}

We first perform the LSA by full-diagonalization of the Hamiltonian $H$ in 
Eq.(\ref{Creutz}), 
under the periodic boundary condition. 
In the LSA, we employ the usual unfolding analysis \cite{Santos}. 
In the unfolding method \cite{Takaishi}, we first prepare a set of 
energy-eigenvalue spectrum $\{E_{i} \}$ 
($i=1,2,\cdots, N_{D}$;  $N_{D}$ is the Hilbert space dimension) in ascending order, 
and then calculate the average level spacing of the original spectrum 
$\{E_{i} \}$ such as $\Delta E=(N_{D}-1)^{-1}(E_{N_{D}}-E_{1})$. 
By using $\Delta E$, we define a new level spacing set $\{s_{i} \}$
as $s_{i}=(E_{i+1}-E_{i})/\Delta E$.
From the set $\{s_{i} \}$, we obtain the statistical distribution $P(s)$, which is
to be compared with the level statistics of the random matrix theory.
When we use multiple realizations (samples) of the disorder, we average $P(s)$
with respect to them to obtain the final result of $P(s)$.

On performing the LSA for the disorder-free case ($\mu=0$), it is important
to note that the system has the translational symmetry. 
This symmetry generally leads to numerous degeneracies in the energy eigenvalues. 
Because of the degeneracies, it is not simple to obtain the probability distributions 
of the level spacing without ambiguities \cite{Nieuwenburg,Schulz}. 
To avoid this difficulty, we consider the cases with small but finite disorders. 
In the presence of disorders, even those that are extremely weak,
the degeneracies of the energy eigenvalues are solved.
In practical calculations, we consider the upper and lower chains with length
$L=16$ and number of particle $N=4$ \cite{numerics1}. 
From the LSA, one can examine the localization properties of the system. 
In general, for an ensemble of localized states, the probability distribution exhibits
Poisson statistics, such as $P_{P}(s)\propto\exp(-s)$, where $s$ denotes the unfolded
level spacing. 
Contrastingly, for an ensemble of delocalized (extended) states, 
the probability distribution is to be GUE, with characteristics such as 
$P_{G}(s)\propto s^2\exp(-4s^{2}/\pi)$
\cite{RM1,RM2,RM3,RM4,RM5,Alet,Oganesyan,Shukle1,Shukle2,GOE}.

Figure~\ref{Fig2} (a) shows the obtained probability distribution for various disorder
strengths for the non-interacting flat-band ($V=0$, $t_1=t_0$).
We find that for a weak disorder ($\mu=1$), the probability distribution is extremely 
similar to Poisson statistics. 
This result indicates the existence of localized states even in a weak disorder.
With increasing disorder strength, we observe an interesting phenomenon, i.e., 
first, the statistics changes from Poisson to GUE-like, and then it
returns to the Poisson statistics.
Calculations for $\mu=6$ and $\mu=30$ shown in Fig.~\ref{Fig2} (a) clearly exhibit
this behavior: Poisson$\to$GUE$\to$Poisson. 
The above behavior of the Creutz ladder model is similar to that in other 
flat-band models in \cite{Shukle1,Shukle2,Goda}. 
The previous studies focus on a single-particle spectrum, however the Creutz model here includes interaction.
The novelty of the results in Fig.~\ref{Fig2} is that {\em even for interacting many-body cases, 
the level statisticsl changes first from Poisson to GUE-like, and return to the Poisson.}
We understand our findings as follows.
The Poisson statistics for the $\mu=30$ ensemble originates from the conventional AL that is induced by disorder.
Contrastingly, the Poisson-like statistics for the $\mu=1$ ensemble 
arises from the flat-band properties of the model.
Crossover takes place from the flat-band localization to
the disorder-induced AL as the disorder increases~\cite{PKI}.
This conclusion is in good agreement with the observation in Sec.~\ref{model}
and will be corroborated by the subsequent IPR calculation.

Figure~\ref{Fig2} (b) shows the LSA of the interacting cases with 
a weak disorder, $\mu=1$.
We find that even for finite interactions $V=1$ and $6$, the Poisson-like 
statistics persists.
This result is indicative of the disorder-free MBL induced by the flat-band structure.
This is again in good agreement with the observation in Sec.~\ref{model}

We also study the non-flat-band case ($t=6t_0$), 
which we regard as a reference system with respect to the AL in finite-size
systems.
Figure.~\ref{Fig2} (c) shows the LSA of a non-interacting non-flat-band
for various $\mu$'s.
The $\mu=1$ and $\mu=6$ results are close to GUE, 
whereas for a larger disorder, $\mu=60$, the conventional disorder-induced 
AL occurs. 
This delocalization-like behavior is robust to the interaction, as shown in
Fig.~\ref{Fig2} (d). 
The obtained result, in particular for the non-interacting case,
seems to contradict the common belief that all the states are localized
in 1D random-potential systems.
Probably, this is a finite-size effect, i.e., for a weak disorder, 
$\mu=1$, localization lengths 
of certain part of states are larger than the system size.
By comparing the results in Figs.~\ref{Fig2} (a) and (b) with those in 
Figs.~\ref{Fig2} (c) and (d),
we find that the localization in the flat-band case is obviously stronger than that in 
the non-flat-band case,
indicating that their mechanisms are different as we discussed in Sec.~\ref{model}.
We will confirm this observation by calculating other quantities. 
The level-spacing ratio in separate energy sectors 
is numerically studied in Sec.~\ref{LSR} to complement the above LSA. 
In addition, we investigate finite-size effects for the LSA in Fig.~\ref{Fig2} (b). 
It is displayed in appendix A.

\subsection{Averaged level spacing ratio}\label{LSR}

\begin{figure}[t]
\begin{center} 
\includegraphics[width=10cm]{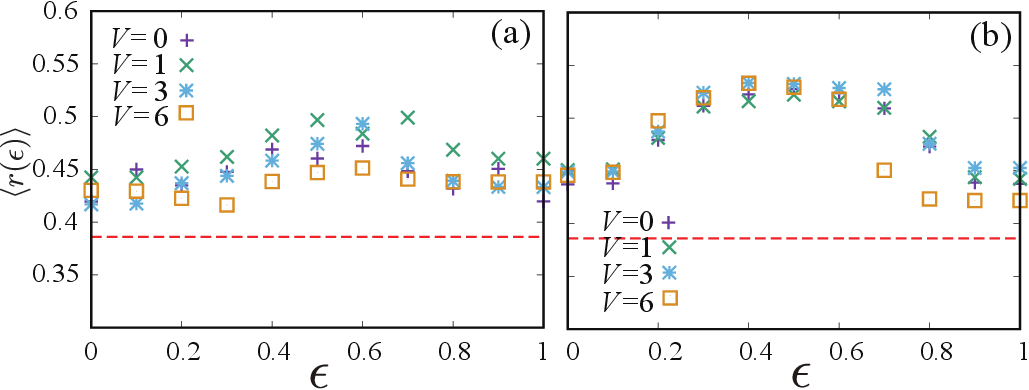} 
\end{center} 
\caption{$V$ dependence of $\langle r(\epsilon)\rangle$: 
The red dotted line represents $\langle r\rangle \sim  0.386$, 
corresponding the ideal value for the Poisson random matrix ensemble, whereas
$\langle r \rangle=0.6$ for the GUE.
$L=16$ and $N=4$ (filling 1/8 case).
(a) for $\mu=1$ and (b) for $\mu=6$.
For all data, we averaged over 20 disorder realizations with different disorder distributions of 
$\mu_{a(b),j}$.}
\label{LSA_r}
\end{figure}
The level spacing ratio (LSR) is often used for study of localization, which is a
kind of numerical analysis of the LSA~\cite{Oganesyan,Luitz}.
In this section, we study the {\em energy-resolved LSR} to see the localization
tendency of various energy  sectors.
To this end, we introduce a normalized energy scale $\epsilon_{i}$, which is defined by $\epsilon_{i}=(E_{i}-E_{N_D})/(E_{1}-E_{N_{D}})$, where $E_{1}$ and $E_{N_D}$ are 
the ground state and maximum excitation energies as before. 
By definition, $0\leq \epsilon_{i} \leq 1$.
LSRs of the energy eigenvalues $\{E_{i} \}$ (in ascending order) are defined as
$r^{k}=[{\rm min}(\delta^{(k)}, \delta^{(k+1)})]/[{\rm max}(\delta^{(k)},\delta^{(k+1)})]$, 
where $\delta^{(k)}=E_{k+1}-E_{k}$. 
To obtain average value $\langle r(\epsilon)\rangle$ as a function of $\epsilon$, 
we average $r^k$ over 1000 energy eigenstates in the vicinity of $\epsilon$ and 20 disorder realizations.
The value of $\langle r(\epsilon)\rangle$ gives us an estimate of the (non-)localization
 tendency of the states around the energy density $\epsilon$. 
For the Poisson random matrix ensemble (localized state), $\langle r \rangle \sim  0.386$. 
On the other hand, for an ergodic state (extended state), 
$\langle r\rangle \sim 0.600$ (GUE).
As we show, $\langle r(\epsilon)\rangle$ in the present system varies from 
$0.4$ to $0.55$.
This result indicates that coexistence of extended and localized states is realized.

For the flat band case ($t_{0}=t_{1}$) in Fig.~\ref{LSA_r}, we display 
$V$-dependence of $\langle r(\epsilon)\rangle$ with the strength of the disorder 
$\mu=1$ and $\mu=6$.
Let us see $V=0,\mu=1$ data first.
All $\langle r(\epsilon)\rangle$s are close to the value of the Poisson distribution 
($\sim  0.386$), but in the intermediate energy region ($\epsilon \sim 0.6$), 
the upward deviation from the Poisson distribution exists. 
This tendency increases for the weak interaction $V=1$, whereas in
the larger interaction cases $V=3$ and $6$, the tendency is weakened.
Therefore, even though there is a small $\epsilon$-dependence in 
$\langle r(\epsilon)\rangle$, the whole states tend to localize 
in the weak disorder and flat-band case. 
This result supports the result in Fig.~\ref{Fig2} in Sec.~\ref{LSA}.
In passing, $V$-dependence in  $\langle r(\epsilon)\rangle$
in Fig.~\ref{LSA_r} (a) may imply a $V$-induced weak spectral 
transition~\cite{Huse}.

Let us turn to the $\mu=6$ case in Fig.~\ref{LSA_r} (b).
It is obvious that $\langle r(\epsilon)\rangle$ has larger values in all cases compared 
with the $\mu=1$ case.
Maximum value of $\langle r(\epsilon)\rangle$ is $0.55$, which is close to the GUE
value.
Therefore, we expect that extended states exist in the region of $\mu=6$, and 
they are located in the center of the energy spectrum.
This observation is in good agreement with the studies of the IPR and the dynamical
behavior of the Creutz ladder given in the subsequent sections.

\subsection{Inverse participation ratio and crossover}\label{IPR}

\begin{figure}[t]
\centering
\includegraphics[width=11cm]{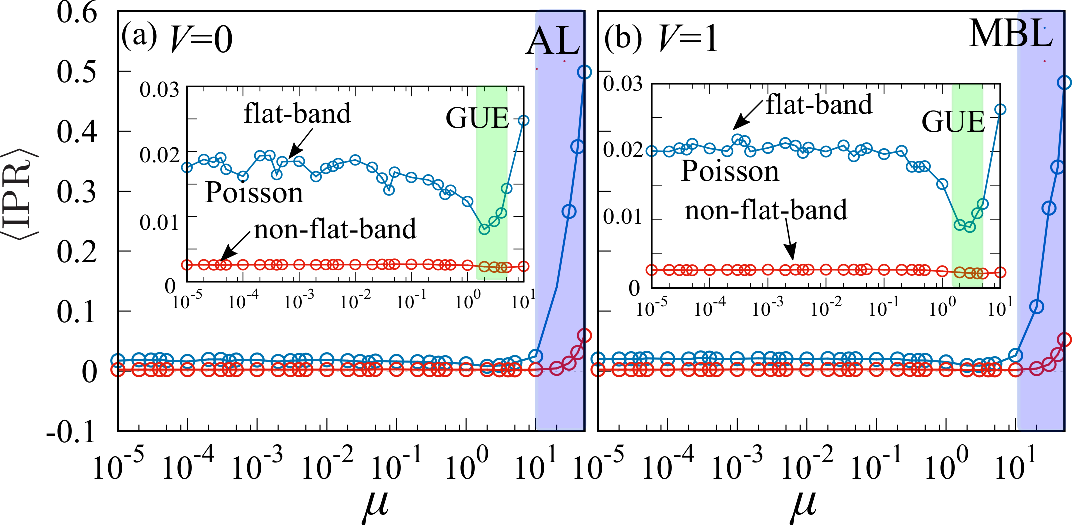}
\caption{Averaged {\rm IPR}: (a) Non-interacting case. 
(b) Interacting case. 
For both the cases, we averaged over 20 disorder samples. 
In both cases (a) and (b), the difference becomes small for $1\lesssim\mu\lesssim 10$, 
where the LSA of the flat-band exhibits GUE-like distribution. 
In the $\mu \gtrsim 10$ regime, the conventional disorder-induced AL/MBL phase appears.
The system size is $L=12$, and the particle number is $N=3$. 
}
\label{Fig4}
\end{figure}

We calculate the IPR, which is often used for the study of localization. 
By diagonalizing the Hamiltonian in Eq.~(\ref{Creutz}),
we obtain all the eigenvectors, 
$|\psi_{\ell}\rangle =\sum_{m}c^{\ell}_{m}|F_{m}\rangle$, 
where $\ell$ labels the eigenstates, $|F_{m}\rangle$ is the Fock-state base and the normalization condition is satisfied $\sum_{m}|c^{\ell}_{m}|^2=1$. 
For these eigenstates, the IPR is defined as $({\rm IPR})_{\ell}=\sum_{m}|c^{\ell}_{m}|^{4}$. 
In particular for the AL with $N$ particles, 
the localization length, $R_\ell$ [in units of the lattice spacing] is given by $({\rm IPR})_{\ell}\simeq 1/(R_\ell)^N$~\cite{Rl}. 
We average $({\rm IPR})_{\ell}$ over all the states for fixed $\mu$ and $V$. 
The averaged IPR is denoted by $\langle {\rm IPR}\rangle$. 

Figure~\ref{Fig4} (a) shows the $\mu$-dependence of $\langle {\rm IPR}\rangle$
in the non-interacting case ($V=0$).
For a sufficiently weak disorder ($\mu \lesssim 1$), the obtained 
$\langle {\rm IPR}\rangle$ in both the flat-band ($t_0=t_1$) and non-flat
band ($t_0=6t_1$) is small compared with that in the strong-disorder regime 
($\mu \gtrsim 10$), where the value of $\langle {\rm IPR}\rangle$ is large 
owing to the existence of the conventional disorder-induced AL. 
In the weak-disorder regime, there exists a clear difference in the $\langle {\rm IPR}\rangle$
of the flat-band and non-flat-band cases~\cite{IPRsmall}, i.e., 
the value of the $\langle {\rm IPR}\rangle$ of the flat-band is 
obviously much larger than that of the non-flat-band, as shown in the inset of
Fig.~\ref{Fig4} (a). 
This means that the flat-band system tends to localize more strongly
than the non-flat-band system~\cite{IPR_length}.
The origin of this difference is clearly explained in Sec.~\ref{model}.
It is intriguing to see that $\langle {\rm IPR}\rangle\simeq 0.02$
gives an estimation of the localization length, $R_\ell\simeq 4.0$,
which is close to the estimation of the localization length given in Sec.~\ref{model}.

It is interesting to observe that in the vicinity $\mu\sim 6$,
$\langle {\rm IPR}\rangle$ decreases in the flat-band system, as shown 
in the inset of Fig.~\ref{Fig4} (a).
This behavior is in good agreement with the results of the LSA presented 
in Fig.~\ref{Fig2} (a) and the LSR in Fig.~\ref{LSA_r}. 
In fact for $\mu=6$, the LSA of the flat-band shows a GUE-like behavior.
Again this behavior of $\langle {\rm IPR}\rangle$ is an evidence of the crossover,
and we estimate $\mu_c\sim 6$.

As our main concern is the MBL state in the flat-band, we study the interacting cases
with finite $V$'s.
Calculations of the IPR for the case, $V=1$, are shown in Fig.~\ref{Fig2} (b). 
We find that the value of $\langle {\rm IPR}\rangle$ of the flat-band increases in 
the weak-disorder regime compared with the $V=0$ case, and it
again decreases considerably near $\mu\sim 6$ as in the $V=0$ case.
We investigated cases for other values of $V$ and found similar behavior
of $\langle {\rm IPR}\rangle$.
We therefore conclude that {\em MBL exists in the flat-band Creutz
ladder model in the weak-disorder regime, reflecting the flat-band structure. 
Moreover, a crossover from flat-band MBL to disorder-induced MBL takes place
as the disorder increases}.
This is one of the main conclusions of this work. 
In Sec.~\ref{DLL}, we shall give a physical picture of the above crossover
that is obtained by calculating energy-resolved localization lengths.

\begin{figure*}[t]
\begin{center} 
\includegraphics[width=17cm]{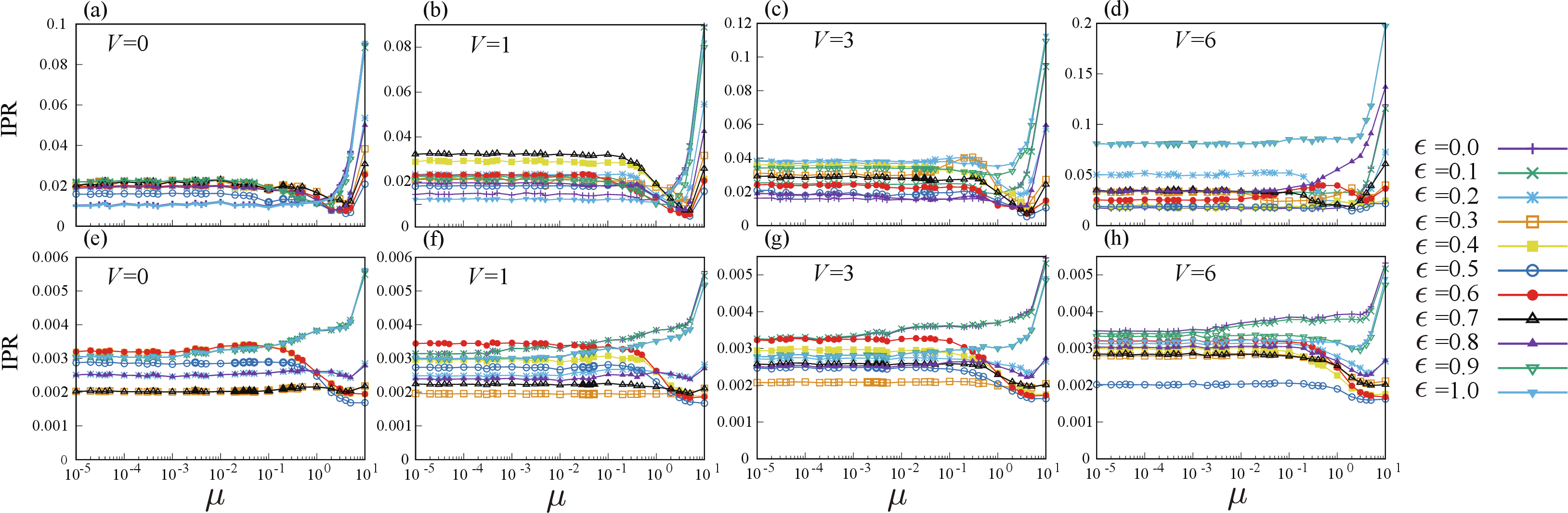}
\end{center} 
\caption{The energy-dependent IPR in the flat band case with 
$V=0\:(a), \:1\:(b),\: 3\:(c),\: 6\:(d)$. 
The IPR in the non-flat band case with $V=0\:(e), \:1\:(f),\: 3\:(g),\: 6\:(h)$.
For all data, $L=12$ and $N=3$ (filling 1/8 case).}  
\label{IPR_energy_density}
\end{figure*}

\subsection{Detailed study of IPR: energy-resolved analysis}\label{IPR2}

In Fig.~\ref{Fig4}, we showed the mean value of the IPR obtained
by averaging all eigenstates. 
We observed that
the IPR exhibits a very interesting behavior as a function of the disorder 
strength $\mu$,
i.e., it substantially decreases in the region $\mu=1.0 \sim 10$.
In Sec.~\ref{IPR}, we emphasized that this behavior of the IPR is consistent with
the LSA and LSR.
In this subsection, we investigate the energy dependence of 
the IPR, $({\rm IPR})_{\ell}$, as we studied the energy-resolved LSR 
$\langle r(\epsilon)\rangle$ in Sec.~\ref{LSR}.
We also study effects of the interactions.

Figure~\ref{IPR_energy_density} shows the disorder $(\mu)$ and interaction $(V)$ dependence of the IPR for states with various energies. 
Results of the flat-band cases ($t_{1}=t_{0}$) are in Figs.~\ref{IPR_energy_density} (a)-(d). 
There, for all $V$s except for $V=6$, the IPR decreases in the region 
$1\lesssim \mu \lesssim 10$ in all energy eigenstates.
In particular, in the central region of $\epsilon$, this behavior is remarkable. 
This indicates that all states tend to extend in the region 
$1\lesssim \mu \lesssim 10$ in the flat-band system.
We think that this peculiar behavior (see the results of the non-flat-band case
below) stems from the fact that in ``weak disorder" below $\mu\simeq 1$, 
all the states sustain properties of the flat-band localization although energy
splitting takes place as a result of the on-site disorder.
In other words for ``strong disorder" ($\mu>10$),  genuine localization due to disorder
takes place as the disorder is strong enough to dominate the flat-band effects.
Therefore, a crossover takes place in the intermediate regime 
$1\lesssim \mu \lesssim 10$, as we explained in the previous sections.

By close look at $V=3$ case in Fig.~\ref{IPR_energy_density} (c), we find that
the data for $\epsilon=0, 0.1, 0.8, 0.9$ and $1.0$ (i.e., areas of the tail of the 
energy spectrum) exhibit only a slight decrease in the IPR in 
$1\lesssim \mu \lesssim 10$. 
This tendency is stronger for the $V=6$ case in Fig.~\ref{IPR_energy_density} (d). 
There, the data for $\epsilon=0, 0.1, 0.8, 0.9$ and $1.0$ shows almost no decrease 
in the value of the IPR in $1\lesssim \mu \lesssim 10$. 
Accordingly, a ``quasi-mobility edge" seems to exist in for $V \gtrsim 3$.

In summary, the IPR of the flat-band regime shows that
for small $V$, as increasing the disorder $\mu$ from the flat-band localization, 
there exists a crossover regime (in $1\lesssim \mu \lesssim 10$) from the flat-band 
localization to the disorder-induced genuine MBL. 
In this crossover regime, all states tend to extend, 
and for larger $\mu$, all states are strongly localized. 
On the other hand for large $V$, such a crossover is blown away, and
the direct transition from the flat-band localization to the disorder-induced 
MBL takes place.
What states are realized in the crossover regime is an interesting problem.
Coexistence of localized and extended states may occur there as  
$\langle r(\epsilon)\rangle$ implies.
It is also important to study if the above properties of the Creutz ladder are
common to other flat-band systems.
These are future works.

We also studied the non-flat band case ($t_{1}=6t_{0}$).
Obtained results of the energy-resolved IPR are shown in
Figs.~\ref{IPR_energy_density} (e)-(h).
For all $V$s, the IPR for $\mu \lesssim 1$ is much smaller than the IPR
of the flat-band case in Figs.~\ref{IPR_energy_density} (a)-(d). 
This result is consistent to the result in Fig.~\ref{Fig4}.
The behaviors of the IPR for each $\epsilon$ in $1\lesssim \mu \lesssim 10$ are almost the same with the different $V$s. 
Furthermore contrary to the flat-band case, states only located in the central region of
the energy spectrum tend to extend and low and especially high-energy states 
tend to localize there.
This behavior comes from the fact that in the non-flat case, quantum states have
different features with each other depending on their energy.
Close look at the data reveals that 
in the weak disorder regime $10^{-2}\lesssim \mu \lesssim 1.0$, the band-edge
states (low and high energy states) start to localize.
This is a common picture of the weak localization.
Data seem to indicate that there exists a transition from the weak disorder to
strong disorder as $\mu$ is increased.
However, location of this transition may depend on each state. 
This behavior is similar to that in the conventional MBL transition in a typical random Heisenberg spin chain \cite{Luitz}.

\subsection{Distribution of localization length and emergence of 
``critical edge''}\label{DLL}

In the pervious two subsections, we calculated the IPRs to study AL and MBL by varying the strength of
the disorder.
The results showed the sharp contrast between the flat and non-flat cases.
In this subsection, we study the distribution of the localization length as a function of energy 
by using the relation between the IPR and localization length, $(IPR)_\ell\simeq 1/(R_\ell)^N$.
This investigation is important to examine the finite-size effect, and to verify that the results of IPR
obtained in the previous subsections for $L=12$ are reliable.
To this end, we study the systems with $L=12$ ($N=3$, 24 sites) and $L=16$ ($N=4$, 32 sites)
focusing on some interesting disorder strengths, $\mu$'s.
Besides the finite-size effect, this investigation reveals very important properties of 
the present system, as we see later in the present subsection.

\begin{figure}[t]
\centering
\includegraphics[width=10cm]{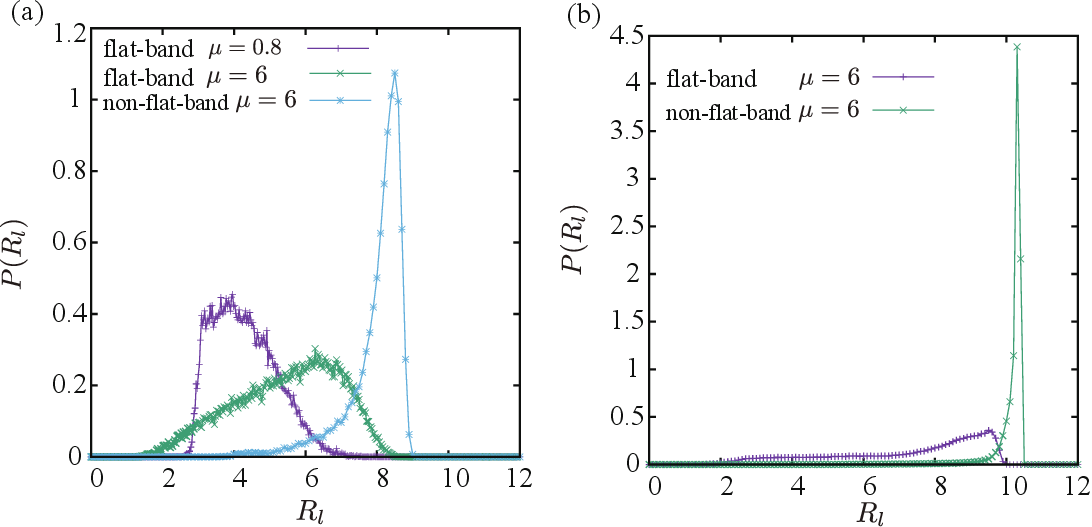} 
\caption{(a) Distribution of localization length, $R_\ell$, for 
the flat-band cases with $V=1, \mu=0.8$ and $V=1, \mu=6$, and 
the nonflat-band case with $V=1, \mu=6$.
System size $L=12$ (24 sites).
(b) Distribution of localization length for 
the flat-band case with $V=1, \mu=6$, and 
the nonflat-band case with $V=1, \mu=6$.
System size $L=16$ (32 sites).
}
\label{LLD1}
\end{figure}

We first show the distributions of the localization length averaged over the entire energy eigenstates,
which correspond to the IPR in Fig.~\ref{Fig4}.
We consider the flat-band cases with $V=1, \mu=0.8$ and $V=1, \mu=6$, and 
the non-flat-band case with $V=1, \mu=6$.
The results for the system size $L=12$ are displayed in Fig.~\ref{LLD1} (a).
For the flat-band case, the localization length for $\mu=6$ is smaller than that for $\mu=0.8$,
which agrees with the calculations of the IPR in Fig.~\ref{Fig4}.
More important observation is that the majority of the localization lengths in the distribution 
in both cases are fairly small compared with the system size, i.e., $\{R_\ell\} <9$.
This result seems to indicate that the system size $L=12$ is large enough to calculate
the localization length for the flat-band case. 
On the other hand for the non-flat band case, typical localization length $R_\ell \sim 9$,
and therefore, the localization length may not be estimated correctly.

To examine the above observation for the flat and non-flat cases, we studied the $L=16$ system.
Obtained results are shown in Fig.~\ref{LLD1} (b).
For the flat-band case, the maximum of the localization length  
$\mbox{Max}\{R_\ell\} \simeq 9$, which is the same with that in the $L=12$ case.
For the non-flat band case, the the maximum of the localization length is slightly larger
than that in the $L=12$ case, but $\mbox{Max}\{R_\ell\} \simeq 10.5$.
These results seem to indicate that the estimations of the localization length are reliable
for both the flat and non-flat cases with the above parameters.

\begin{figure}[t]
\centering
\includegraphics[width=10cm]{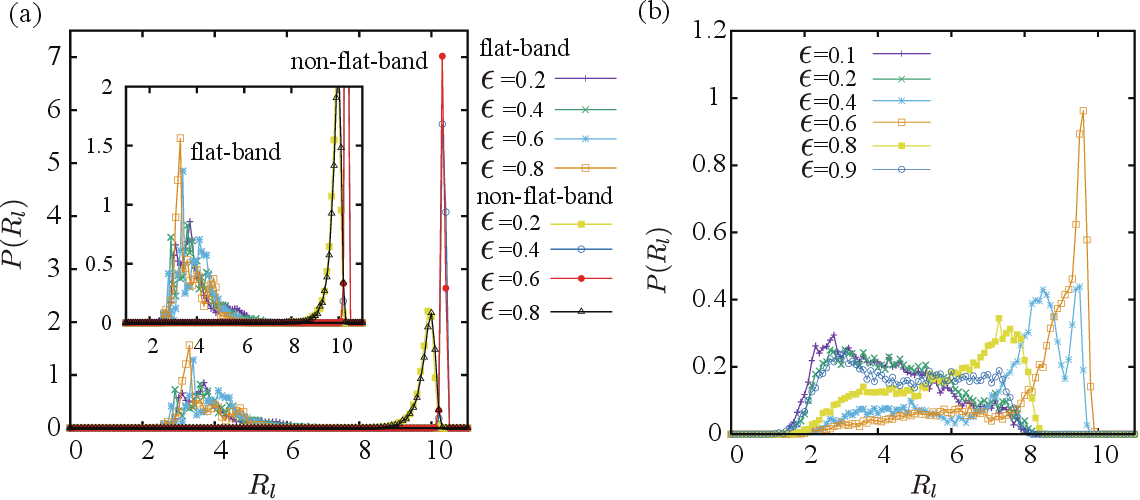} 
\caption{
(a) Energy-resolved distribution of localization length, $R_\ell$, for 
the flat-band case with $V=1, \mu=0.1$, and 
the nonflat-band case with $V=1, \mu=6$.
System size $L=16$ (32 sites).
(b) Energy-resolved distribution of localization length for 
the flat-band cases with $V=1, \mu=6$.
System size $L=16$ (32 sites).
Distribution exhibits a similar shape to that of the flat-band case (non-flat band case)
in the edge regimes (central regime) of the energy  spectrum.
5 realizations of disorder and $10^4$ eigenstates (in the very vicinity of $\epsilon$) are used for each energy $\epsilon$.}
\label{LLD2}
\end{figure}

We also studied the energy-resolved localization length for the $L=16$ system with the above parameters
and obtained very important observations.
The calculations are shown in Fig.~\ref{LLD2} (a) and (b).
For the non-flat-band system in  Fig.~\ref{LLD2} (a), the distribution is dominated by a sharp peak and 
there is a moderate peak very close to the sharp peak. 
For the flat-band case with a small chemical potential $\mu=0.1$, the distribution has a single moderate
peak centered at $R_\ell=3$.

\begin{figure}[t]
\centering
\includegraphics[width=10cm]{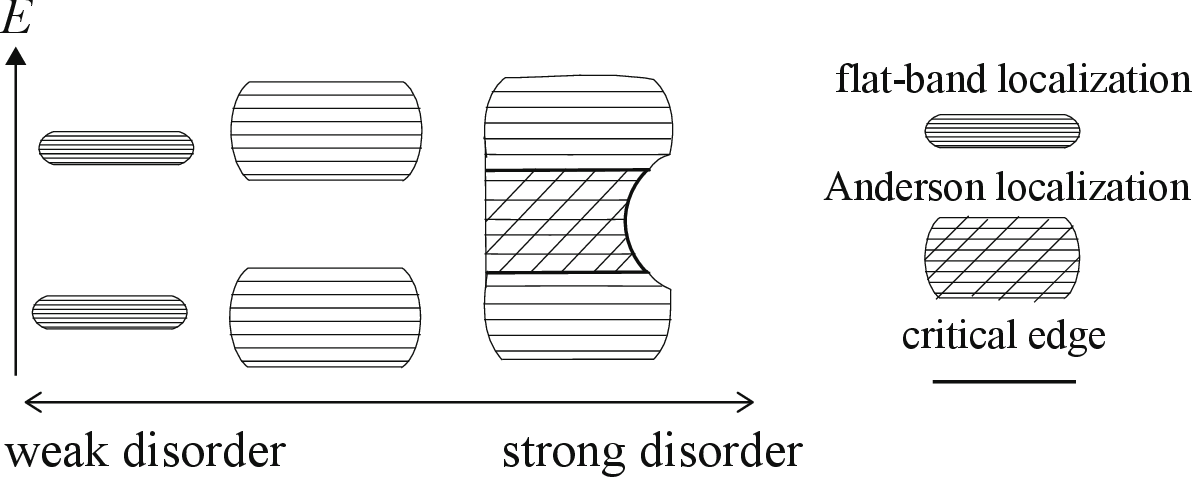} 
\caption{
Schematic picture of localization and band structure as a function of the strength of disorder.
Mechanism of MBL in the interaction systems changes from the flat-band localization to
the Anderson-like localization as the strength of disorder increases.
For suffiecently large $\mu$, the flat-band structure is destroyed and two bands merge.
At the central regime in the energy spectrum, localization similar to AL takes place and 
the localization length in that regime is larger than that at the edges of the energy spectrum. 
}
\label{LLD3}
\end{figure}

On the other hand for the flat-band case with $\mu=6$ in Fig.~\ref{LLD2} (b), 
the distribution has a different shape depending on eigenenergy.
States in the band center have a fairely large localization length, whereas at 
the band edges, the states are localized.
From the calculations of the non-flat band case with $\mu=6$ and flat-band case with $\mu=0.1$ in 
Fig.~\ref{LLD2} (a), 
we observe that the states far from the band center are localized due to the flat-band localization,
which is one of the properties of the genuine Creutz ladder system. 
On the other hand for the states in the band-center regime, AL caused by the disorder potential
is the main mechanism of localization as in the non-flat system.
In other words, there exists a critical strength of the disorder at which the flat-band structure
shown in Fig.~\ref{Setup}(b) is destroyed, and the upper and lower bands merge.
In this sense, there exists a ``critical edges'' separating the flat-band localized and  
AL regimes.
[We estimate them as $\epsilon=0.3$ and $\epsilon=0.85$, respectively.]
Schematic picture is shown in Fig.~\ref{LLD3}, which displays an intuitive understanding of 
the crossover observed by the LSA and IPR.
Anyway, more detailed study on this kind of crossover is a future problem.

\subsection{Ergodicity-breaking dynamics}\label{dynamics}

The above results of the LSA and IPR indicate that disorder-free 
single-particle localization and MBL occur in the flat-band Creutz ladder. 
This motivates us to simulate the dynamics of the Creutz ladder. 
In the conventional disorder-induced AL and MBL, information of an initial density wave pattern 
is stored for long times \cite{Nandkishore,Basko,Abanin,Alet,Abanin2}.
This behavior is a hallmark of ergodicity breaking and indicates the breaking of
the eigenstate thermalization hypothesis~\cite{Nandkishore,Basko,Abanin,Alet,Abanin2,Sierant}.
Here, we focus on the disorder-free cases and investigate
whether the flat-band Creutz ladder exhibits ergodicity breaking dynamics. 
To this end, we employ the time-dependent exact diagonalization method 
with the periodic boundary condition~\cite{Prelovsek,Manmana}.

As discussed in Sec.\ref{model}, a particle wave function in the flat-band regime 
tends to have a non-vanishing amplitude only on definite adjacent finite sites.
Therefore, we expect that the localization of the flat-band system
exhibits different behavior depending on the particle fillings.
This expectation obviously comes from the observation that
the Pauli exclusion principle and the repulsion work substantially
at large fillings but less effectively at low fillings.
In fact at large fillings, the repulsions between particles come to effective, and 
they suppress movements of the particles. 
As a result, localization is enhanced.

\begin{figure}[t]
\centering
\includegraphics[width=7cm]{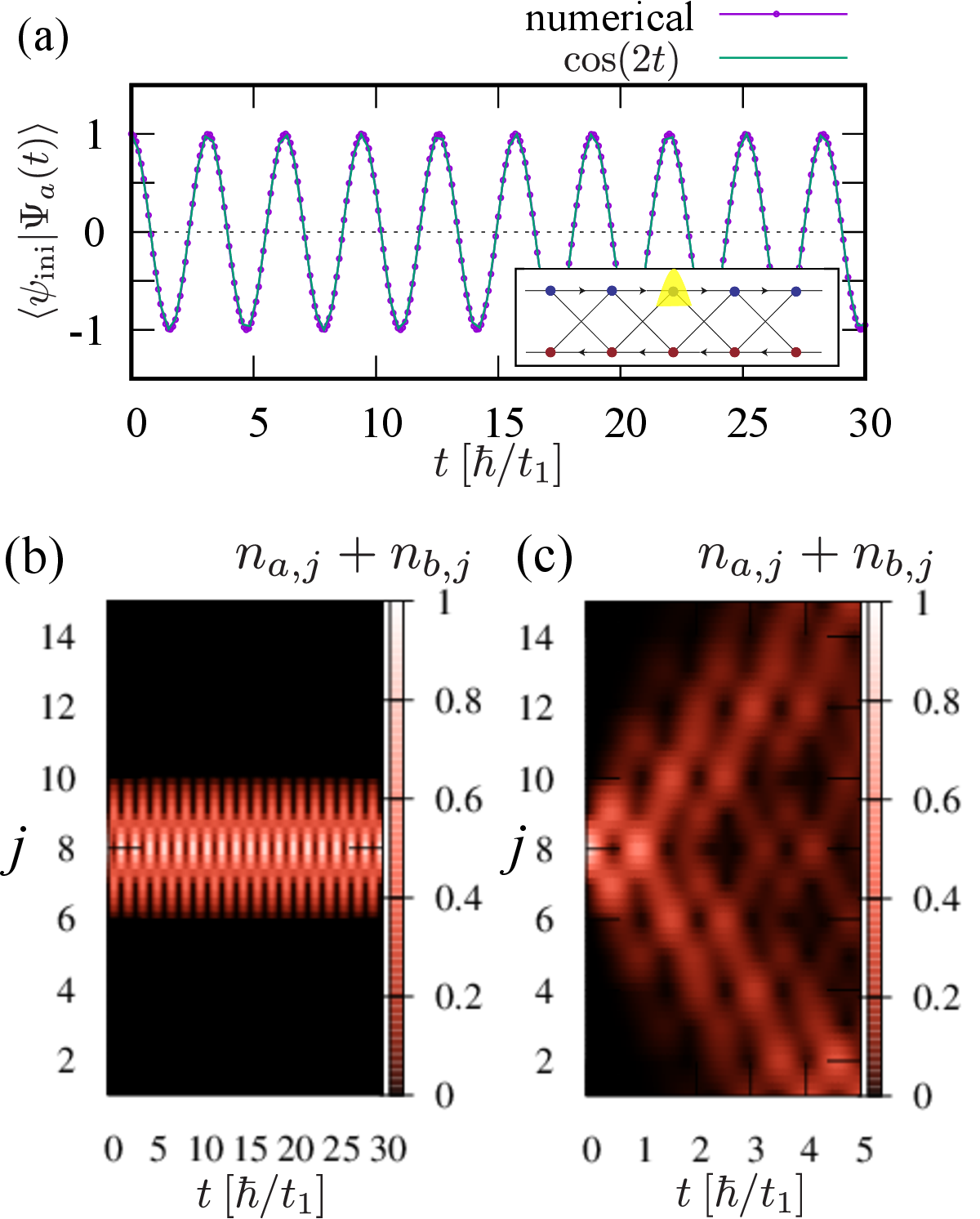}
\caption{One particle localization dynamics. (a) Return probability amplitude: The purple line data represents the numerical time evolution of $\langle \psi_{\rm ini}|\Psi_{a}(t)\rangle$ 
with the initial state $|\psi_{\rm ini}\rangle =a^{\dagger}_{j=8}|0\rangle$.
The green line is the analytical oscillation solution of $\langle \psi_{\rm ini}|\Psi_{a}(t)\rangle$ obtained by Eq. (\ref{solution}).
Density distribution dynamics for the rung, $n_{a,j}+n_{b,j}$
(b) for the flat-band condition, (c) for the non-flat band condition, $t_{0}=2t_{1}$.}
\label{Fig9}
\end{figure}

To verify the above expectation, we investigate three cases of particle filling, $1/8$,
$1/6$ and $1/4$-fillings in our numerics. 
To see their dynamics, we prepare a specific initial state for each filling
such as, 
\begin{eqnarray} 
|\psi_{\rm ini}\rangle = \prod^{2qL}_{i=1} a^{\dagger}_{(2q)^{-1}(i-1)+1}|0\rangle, \label{initial}
\end{eqnarray}
where $q$ is taken as follows for each filling, $q=1/8$, $1/6$ and $1/4$,
respectively.
This initial state is a totally non-entangled Fock state, and 
therefore it is quite suitable for detect the localization dynamics \cite{Abanin2}.
To characterize the localization dynamics, we measure the long-time average of 
the return probability \cite{Nieuwenburg,Luca},
\begin{eqnarray}
P = \lim_{t\to \infty}P(t)
=\lim_{t\to \infty}\frac{1}{t}\int^{t}_{0}|\langle \psi_{\rm ini}|
e^{-i\frac{H}{\hbar}t'}|\psi_{\rm ini}\rangle |^2 dt', \label{aveP}
\end{eqnarray}
where $P(t)$ is a return probability at $t$. 
Here, as shown in \cite{Abanin,Nieuwenburg,Luca}, if the initial state is 
given by $|\psi_{\rm ini}\rangle=\sum_{\ell}d_{\ell}|\psi_{\ell}\rangle$
where $d_{\ell}$ is a coefficient of an eigenstate 
$|\psi_{\ell}\rangle$ in the quenched Hamiltonian, 
$P=\sum_{\ell,k}|d_{\ell}|^2|d_{k}|^{2}\delta_{\epsilon_{k},\epsilon_{\ell}}$, where $\epsilon_{\ell}$ 
is the eigenenergy of $|\psi_{\ell}\rangle$. 
Accordingly, $P$ is related to the level spacing of the eigenenergy. 
That is, the above expression of $P$ indicates that states with small level spacings contribute
more to $P$. 
Since the Poisson distribution (realized in localized regimes) has a small level-spacing regime, 
the value of $P$ tends to be large. 
Therefore, localization enhances the value of $P$.
In a conventional localization state, entanglement of eigenstates is fairly suppressed, and
each eigenstate $|\psi_{\ell}\rangle$ tends to be close to the Fock state $|F_{m}\rangle$. 
Then, a finite (IPR)$_\ell$ implies a finite $P$ although
$P$ is indirectly related to IPR defined in Sec.~\ref{IPR}.

If the value of $P$ is finite, the memory of the initial state is preserved. 
This implies that an ergodicity breaking takes place and the system exhibits localization.
In our practical numerics, we set the unit of time by $\hbar/t_{1}$, 
set the long-time limit as $t=10^{3}$[$\hbar/t_{1}$] in Eq.~(\ref{aveP}), 
and use the time slice, $dt=10^{-3}$ [$\hbar/t_{1}$]. 
We put $\mu =0$ for all the calculations.

To begin with, let us verify a single particle localization dynamics. 
The initial state is set to $|\psi_{\rm ini}\rangle =a^{\dagger}_{j=8}|0\rangle$ 
with the system size $L=15$.
Figure~\ref{Fig9} (a) shows the dynamical behavior of 
$\langle \psi_{\rm ini}|e^{-i\frac{H}{\hbar}t}|\psi_{\rm ini}\rangle$. 
The numerical result exhibits a clear localization 
since $\langle \psi_{\rm ini}|e^{-i\frac{H}{\hbar}t}|\psi_{\rm ini}\rangle$ oscillates, 
and also its oscillating period agrees with the analytical result of Eq.~(\ref{solution}). 
Under the flat-band condition $t_{0}=t_{1}$, the single particle certainly localizes. 
The detailed density dynamics for rung $j=8$, $n_{a,j=8}+n_{b,j=8}$ is also plotted 
in Fig.~\ref{Fig9} (b) for the flat-band case and (c) for non-flat band case. 
For the flat band case, the initial single particle is localized with a oscillation between $j=7$ and $j=9$ rungs, 
corresponding to the analytical result of Eq.~(\ref{solution}).
On the other hand, see Fig.~\ref{Fig9} (c) for the non-flat band case $t_{1}\neq t_{0}$ 
with $V=0$, the oscillation of 
$\langle \psi_{\rm ini}|e^{-i\frac{H}{\hbar}t}|\psi_{\rm ini}\rangle$ decays immediately.

\begin{figure}[t]
\centering
\includegraphics[width=9cm]{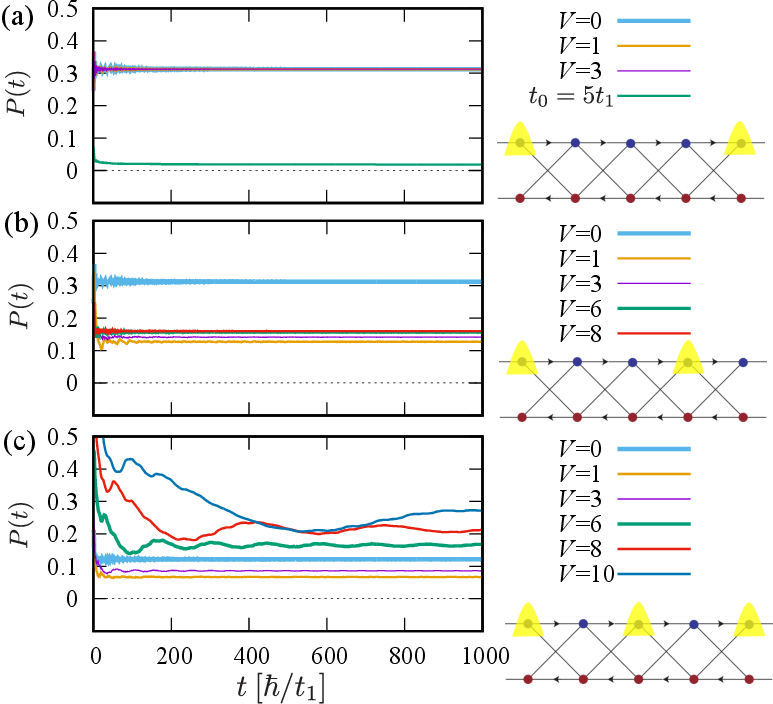}
\caption{Exact dynamics for small particle system: (a) $L=12$ system with 1/8 filling.  
(b) $L=9$ system with 1/6 filling. (c) $L=8$ system with 1/4 filling.}
\label{Fig10}
\end{figure}

Let us turn to the multiple-particle system.
We calculated $P(t)$ for various filling cases and interaction strengths $V$.
First, we consider the $1/8$-filling case.
We expect that the inter-particle distance of the initial state is sufficiently large there,
and the particles do not substantially interact with each other.
However, this does not necessarily mean that the repulsion does not influence
the dynamics of each particle at all.
The numerical result for the $L=12$ three-particle system is displayed in 
Fig.~\ref{Fig10} (a). 
For various $V$s in the flat-band case, $P(t)$ takes a finite large value for long times, 
i.e., $P\sim 0.37$. 
This indicates the strong localization of particles and the ergodicity breaking. 
The independence of the value of $V$ in the dynamics originates from 
the large inter-particle distance. 
On the other hand, the results for the non-flat band case with $t_{0}=2t_{1}$
show that the value of $P(t)$ suddenly decays, and therefore 
the dynamics of the non-flat band system is ergodic. 
These numerical results are consistent with the results of the level spacing analysis 
shown in Fig.~\ref{Fig2} (b). 

Second, let us turn to the $1/6$-filling case. 
For the initial state of Eq.~(\ref{initial}) in the non-interaction case $V=0$, 
we expect that each particle starts to oscillate around the adjacent rungs as
described by the single-particle solution of Eq.~(\ref{solution}) in Sec.~\ref{model}.
There, although each single particle wave function spreads a little, the overlap and
interference between particles are not substantial, 
and therefore we expect the multiple-particle system for $V=0$ exhibits strong localization similar to the $1/8$-filling case above. 
However, once $V$ is switched on, oscillating single particles start to interact 
with each other, and the single-particle localization picture may be affected 
by the existence of the interaction $V$. 
Figure.~\ref{Fig10} (b) is the result of $P(t)$ for the $L=9$ three-particle system. 
For $V=0$, as we have expected, $P(t)$ exhibits strong localization $P\sim 0.37$. 
Remarkably for a finite $V$, $P(t)$ remains a definitely finite value, $P\sim 0.15$.
Even for a finite $V$ with the $1/6$ filling, the system exhibits the ergodicity-breaking
dynamics, but the value of $P$ is a little smaller than that of the $V=0$ case and
the $1/8$ filling, i.e., the interacting system is moderately localized. 
For this moderate-localization regime, it is difficult 
to judge whether the LSA and LSR obey Poisson or GUE ensembles. 
The calculations of the LSA and LSR shown in Appendix B prove this expectation.
Properties of the moderate localization are interesting and warrant deep study 
as a future work. 

Third, we focus on the $1/4$-filling case. 
The inter-particle distance is small and 
the overlap of the single particle oscillating wave functions is so large that 
we expect the interaction $V$ drastically changes the localization properties of 
the system. 
Figure.~\ref{Fig10} (c) is the result of $P(t)$ for the $L=8$ four-particle system.
Interestingly enough, depending on the value of $V$, the dynamical behavior of 
$P(t)$ drastically changes.
For the non-interacting $V=0$, the moderate localization appears since $P\sim 0.13$.
As increasing $V$ from $V=0$, for weak but finite $V$ cases, 
the localization is highly suppressed, i.e., the system tends to be extensive since $P < 0.1$.
However for large $V \gtrsim 6$, the $P$ increases to $P>0.2$.
That is, the interaction $V$ suppresses the localization tendency first, but it starts to
enhance localization as $V$ exceeds a critical value.
We expect that in the localized regime for large $V$, the particles repel 
each other strongly, and then particles are squeezed and localized moderately.
The localization length of the moderate localization for large $V$ may be a little 
larger than that of the strong localization with $P\sim 0.37$.
Calculations of the averaged LSR, $\langle r(\epsilon) \rangle$, in appendix B
suggest that the band-edge eigenstates in the moderate-localized regime
have stronger tendency of delocalization compared with the strong-localized states. 

\begin{figure}[t]
\centering
\includegraphics[width=6.5cm]{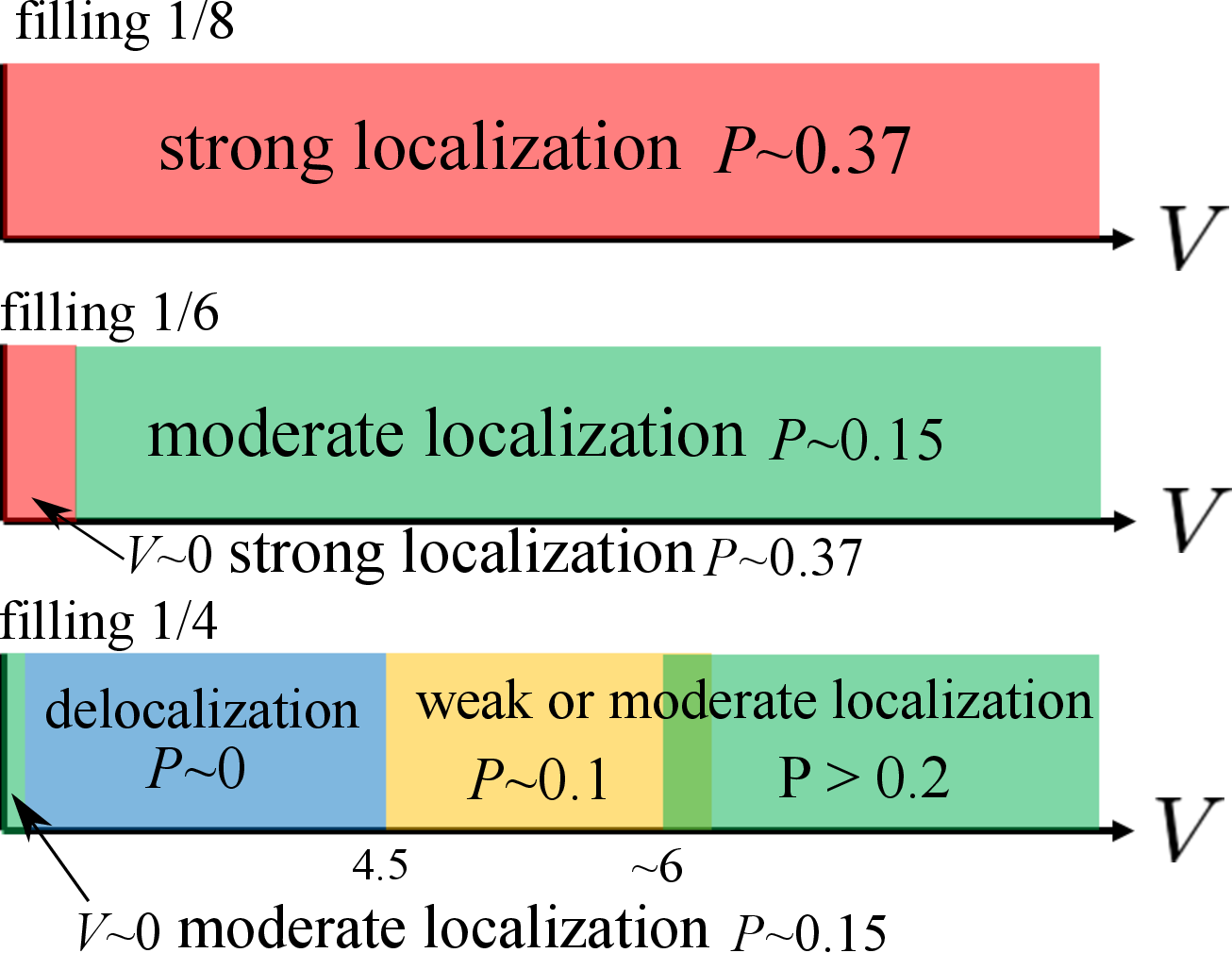}
\caption{Qualitative tendency of localization obtained by the calculation $P(t)$.}  
\label{Fig11}
\end{figure}

We conclude that for small size systems, a disorder-free MBL exists in 
the flat-band Creutz ladder both with and without interactions, and 
{\em it exhibits ergodicity-breaking dynamics}. 
In Fig.~\ref{Fig11}, we summarize the results of the numerical calculations and 
show the qualitative dynamical properties of the system as a function of
the interaction $V$ for various particle fillings.  


\section{Conclusion}\label{conclusion}

In this work, we have clarified disorder-free single-particle localization and 
MBL phenomena induced by the flat-band structure of the Creutz ladder model. 
We found that the flat-band localization originates from a localized Wannier state (FB compacton), 
and the localization length is quite short compared to that of ordinary AL.
The localization length of the flat-band Creutz ladder system
is also insensitive to the strength of the disorder although the localization length of AL
is strongly influeneced by the disorder strength.
As a result, effects of interactions in flat-band localization depends on particle filling 
(inter-particle distance) significantly.

In Sec.~\ref{results}, we extensively studied the flat-band localization properties 
by using some conventional numerical methods.
We extracted the localization properties from the statistical properties of the static spectrum 
and eigenstates in the Creutz ladder Hamiltonian. 
In the flat-band regime, the LSA  exhibits Poisson distribution in the weak-disorder regime 
with or without interactions. 
This indicates that the flat-band model exhibits a (many-body) localization induced by 
the flat-band nature not by disorder as in AL.
After that, we calculated the LSR from the spectrum and also the IPR from eigenstates of the model 
in order to capture the localization tendency in the real space. 
We found that they support the LSA result.

We also studied the flat-band localization from the view point of the dynamical aspect.
We found that flat-band localization tends to prevent the system from thermalization. 
The single particle localization picture was analytically given in Sec.~\ref{model}. 
If we put on a single particle on a single site on the flat-band system, the single particle localizes 
with oscillating. 
To estimate this dynamical localization with or without interactions, we performed exact dynamical 
simulations for small size systems. 
To judge the thermalization and ergodicity breaking, we employed the return probability, 
which quantifies how much information of initial state wave function remains. 
By calculating the long-time average of the return probability, 
we characterized an ergodicity-breaking dynamics similar to the conventional 
disorder-induced AL and MBL dynamics, and also found rich localization properties 
as varying particle filling and the repulsive interaction.
In summary, even in the interacting cases, the system exhibits localization and ergodicity-breaking dynamics. 
Our numerics is only for small system sizes but exact, 
and therefore our results can be a benchmark for future simulations with 
large system sizes, e.g. by using Krylov subspace method. 
We also expect that the findings in the present work are useful
for future real experiments on cold atoms such as \cite{synthetic_dim1,Kang}.

We also expect similar phenomena in other flat-band models, 
such as the saw-tooth and Lieb lattice models, which are to be realized in experiments \cite{Zhang,Taie, Mukheriee,Vicencio}, 
and also some studies in the Creutz ladder in the clean limit \cite{Tovmasyan1,Tovmasyan2} pointed out the presence of conserved quantities. 
Such conjecture may support our results.


\section*{Acknowledgments}
Y. K. acknowledges the support of a Grant-in-Aid for JSPS
Fellows (No.17J00486). 


\section*{Appendix A. System size dependence of LSA}
\begin{figure}[h]
\begin{center} 
\includegraphics[width=8cm]{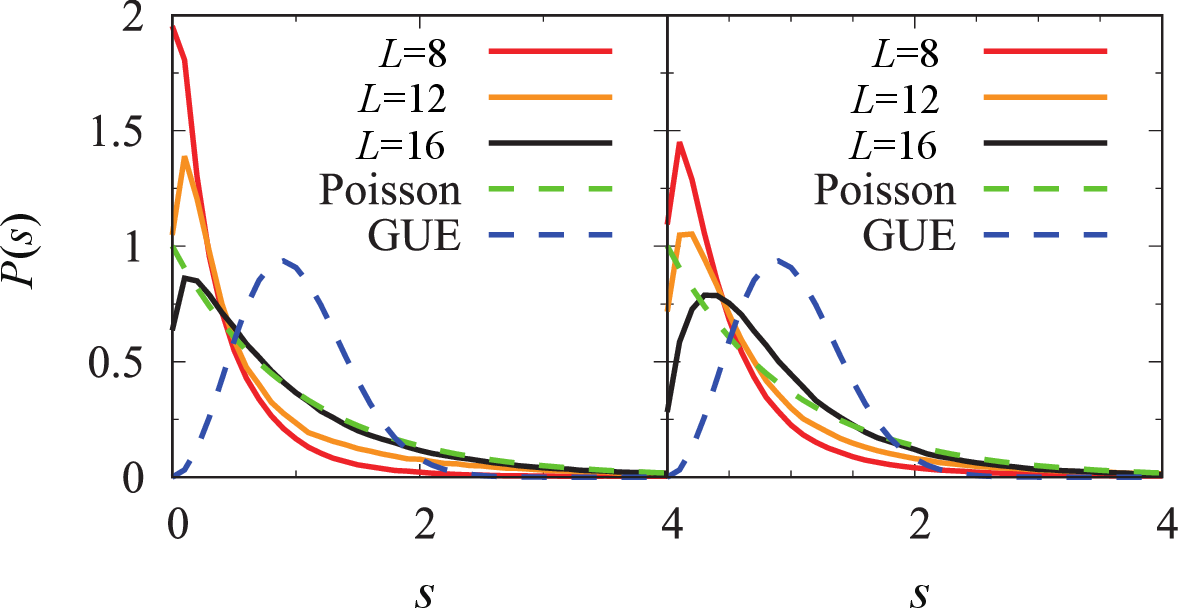}
\end{center} 
\vspace{-0.3cm}
\caption{System-size dependence of the level spacing analysis: (a) non-interacting 
flat-band case with $\mu=1$. (b) interacting flat-band case with $\mu=1$.
In both the cases, the level spacing distribution, $P(s)$, approaches the Poisson
distribution as the system size is increased.}  
\label{sup1}
\end{figure}
We calculate the statistical distribution by using the unfolded level 
spacing method. 
Here, we show its system-size dependence for the non-interacting flat-band case
in Fig.~\ref{sup1} (a).
For the $L=8$ case, the shape of the probability distribution is different from 
that of the Poisson distribution. 
The value of $P(s)$ near $s\sim 0$ tends to increase for a small system size.
On increasing the system size up to $L=16$, the probability 
distribution can be regarded as Poisson-like. 
From this result, we expect that for larger system sizes, 
the probability distribution approaches the exact Poisson distribution. 
Therefore, for the non-interacting flat-band system,
localization can be clearly observed for a large system size. 
Such a system-size dependence is also exhibited for the interacting case. 
Figure~\ref{sup1} (b) shows the system-size dependence of the LSA for the $V=1$ case. 
Compared with the non-interacting case, the increasing tendency of $P(s)$ 
in the vicinity of $s\sim 0$ is weak in small systems. 
However, the probability distribution deviates from the exact Poisson distribution. 
On increasing the system size up to $L=16$, the probability distribution 
approaches the Poisson distribution. 

\section*{Appendix B. Averaged level spacing ratio of 1/6 and 1/4-fillings}
\begin{figure}[h]
\begin{center} 
\includegraphics[width=8cm]{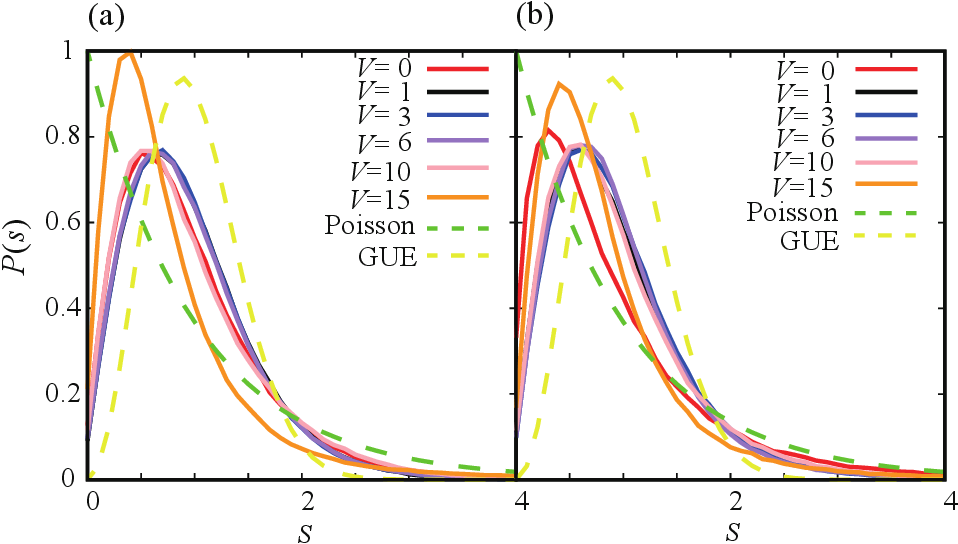}
\end{center} 
\caption{Level-spacing analysis: Interaction dependence in disordered flat-band ($\mu=1$).
(a) 1/4-filling for $L=10$ system. 
(b) 1/6-filling for $L=12$ system.}  
\label{sup2}
\end{figure}
In this appendix, we show the LSA and LSR for the 1/4 and 1/6-filling cases.
The results of the LSA for the 1/4 and 1/6-fillings are shown in Fig.~\ref{sup2} (a)--(b).
Similarly to Fig.~\ref{Fig2} (b), we add the disorder $\mu=1$ in order to avoid the degeneracies. 
For the 1/4-filling, the results from $V=0$ to $V=10$ exhibit the almost same behavior, 
that is, the statistics is neither the Poisson nor GUE distribution.
But for $V=15$, the statistics gets closer to the Poisson distribution.
For the 1/6-filling, the result of $V=0$ is closer to the Poisson than 
the other cases of $V$. 
The results from $V=0$ to $V=10$ exhibit similar behavior, i.e.,
the statistics is neither the Poisson nor GUE distribution. 
But for $V=15$, the statistics gets slightly closer to the Poisson distribution.

\begin{figure}[h]
\begin{center} 
\includegraphics[width=10cm]{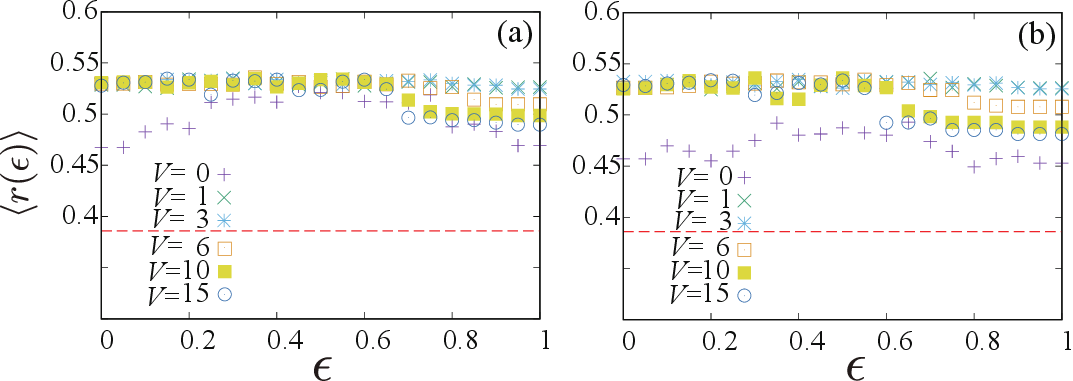} 
\end{center} 
\caption{$V$-dependence of $\langle r(\epsilon)\rangle$ with $\mu=1$:
(a) 1/4-filling case for $L=10$. (b) 1/6-filling case for $L=12$.
The red dotted line represents $\langle r\rangle \sim  0.386$, 
corresponding the ideal value for the Poisson random matrix ensemble, 
whereas $\langle r \rangle=0.6$ for the GUE.}  
\label{sup3}
\end{figure}

The results of the LSR for the flat-band case with 1/4 and 1/6-fillings are shown 
in Fig.~\ref{sup3} (a)--(b).
Similarly to Fig.~\ref{LSA_r} (a), we add the disorder $\mu=1$. 
For both filling cases, $V=0$ results are tend to be delocalized. 
For all finite-$V$ results, the delocalization tendency of higher band-edge eigenstates 
is suppressed, and for larger $V$ the localization tendency of higher band-edge eigenstates seems to increase.



\begin{thebibliography}{10}
\bibitem{diss_come}
The disorders act independently in each site and they effect
destroying the localized Wannier state in Eq.~(\ref{localized_gw}). 
If one employs chiral symmetric disorders instead, the system is fairly robust against 
the disorders.
\bibitem{t16t0}
Such a hopping amplitude ratio can be easily realized in real experiments \cite{Kang}. 
By controlling a modulation amplitude of a driving optical lattice, $t_{0}=t_{1}$ condition can be achieved.
\bibitem{numerics1}
The Hilbert space dimension expanded by the Fock state is $35960$, and 
we discard the top and bottom 10$\%$ of the energy eigenvalues to obtain 
a clear distribution.
\bibitem{GOE}
The probability distribution is not Gaussian orthogonal ensemble 
because our Hamiltonian matrix is not real-symmetric but only Hermitian.
\bibitem{Rl}
More precisely for the AL without inter-particle interactions,
$\langle {\rm IPR}\rangle\simeq \prod^N_{n=1}(1/R_n)$, where $n$ labels quantum states
occupied by $N$-particles, and $R_n$ is the localization length of $n$-th state. See Ref. \cite{Takaishi}.
\bibitem{IPRsmall}
The small values of $\langle {\rm IPR}\rangle$ are partly owing to 
the degeneracy originating from the translational symmetry of the model. 
In the small-disorder regime, the breakdown of the translational symmetry is weak. 
Accordingly, there exist numerous quasi-degenerate states.
\bibitem{IPR_length}
Simple estimation of the average of $R_\ell$, $\langle R \rangle$,
is obtained by using the obtained results of $\langle {\rm IPR}\rangle$.
It gives $\langle R \rangle\simeq 8$ for the non-flat-band and 
$\langle R \rangle \simeq 4$ for the flat-band. 
This result for the flat-band is reminiscent of the Wannier state in Eq.~(\ref{localized_gw}).
\end{thebibliography}
\end{document}